# Assessment Of Selected Trace Elements Concentration in Eleyele Lake-Water, Ibadan South Western Nigeria

**Kunle-John, I. O., Michaels, S. P., and Okay E. N.**

National Water Resources Institute, Kaduna

**Abstract**

Eleyele Lake has enormous economic importance as it is completely surrounded by various communities which discharge their domestic waste directly into the lake. This alters the physical, chemical and biological characteristics of the lake. It is essential to assess the water for its various usage. Twelve (12) samples were collected from various locations of the lake and analysed. Some physical parameters (electrical conductivity, Total Dissolved Solids, Ph and temperature were determined in-situ. The rest of the sample was taken to the laboratory for various chemical analysis and the results were compared to the WHO standards. The chemical extent of the contamination was determined by the contamination factor, degree of contamination and Geo-Accumulation Index. The physical parameters show that the TDS has an average of 122.2ppm and EC was uniform throughout the various points of reading suggesting that the lake is fresh water. The pH averaged at 72, temperature at 27.2 degrees. The selected trace element falls within the WHO acceptable limits. Their contamination indices showed that Ba, Co, Cs, Cu, Mo, Rb, Sr and Zn are generally less than one depicting their geogenic origin. The high degree of contamination is influenced by high levels of Al and Fe due to human activities and industrial waste disposal and can lead to anemia, osteomalacia (brittle or soft bones), cardiac arrest, stomach problems, nausea, and hemochromatosis. Thus, Eleyele lake is not advisable for consumption.

**Key words:** Eleyele lake, Contamination factor, Geo-Accumulation Index

Trace elements concentration,

## INTRODUCTION

Eleyele Lake serves a great economic importance; it provides water for domestic agricultural and industrial use, supporting subsistence and artisanal fisheries. It is completely surrounded by various communities that discharge their domestic waste directly into the lake water. When wastes from different sources are discharged into the water body that alter the physical, chemical and biological characteristics of the water body in such a way that it may not be useful for the purpose for which it is intended. Environmental pollutants arising from anthropogenic source have the potential to affect the aquatic ecosystem in a synergistic manner. The productivity and growth of aquatic organisms depend on the physicochemical characteristics of the water body and maximum productivity can only be obtained at optimal levels of physico-chemical parameters. It is therefore very essential and important to test the water before it is used for drinking, domestic, agricultural or industrial purposes. Quality assessment involves the analysis of physico-chemical, biological and microbiological parameters that reflect the biotic and abiotic status of the ecosystem. Water quality monitoring has a high priority for the determination of current use. The usefulness of water depends on whether such waters are timely, quantitatively and qualitatively available according to Bhatia (2010).

The sources of water for any specific purpose are not as important as the suitability of the water for the desired purpose. With increasing human population, industrialization. urbanization and the consequent increase in demand for water for both domestic and industrial uses, the attendant increase in the implication of polluted water on man and the environment have been severally studied (Asiwaju-Bello and Akande. 2004: Ige *et al*., 2008)

High inputs and presence of trace metals contaminants in water could pose serious problems resulting in potential long-term toxic implication on the environment and the ecosystem. When the main water body of a town is contaminated by heavy metals and trace elements in general, it has a direct effect on that environment and ecosystem.

The aim of the study is to assess the potential impact of selected trace metals found in Eleyele lake on groundwater in parts of Ibadan Southwestern Nigeria.

**Location of the Study Area**

This study was carried out in Eleyele (Figure 1), which is located in the northeastern part of Ibadan city, southwestern Nigeria within longitude 3° 33'E and latitude 7 25 18" N-7 26 48' N. The artificial lake was constructed in 1942 to provide raw water to provide a potable water supply for major parts of Ibadan metropolis. The lake and associated dams at Eleyele receive water from the River Alapata and headstream of River Onal.

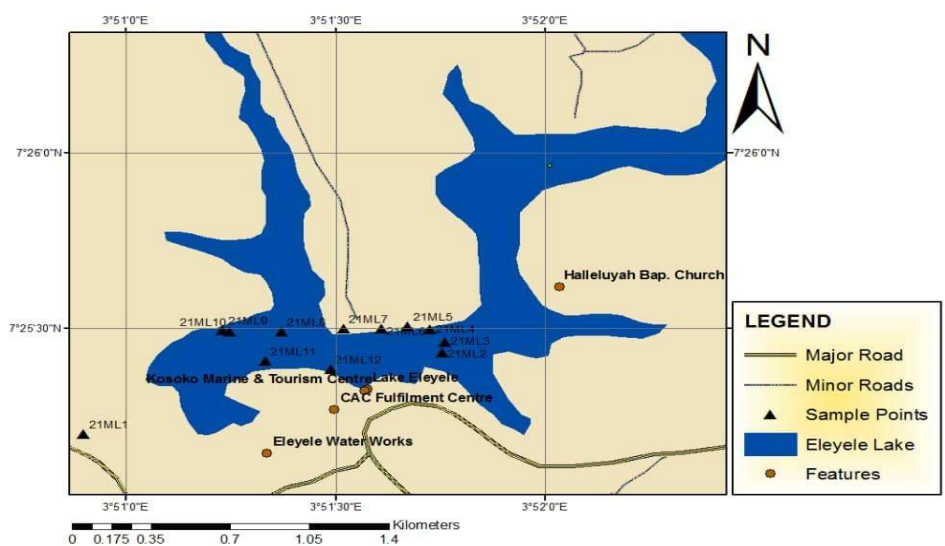

Figure 1**:** Location map of Study Area showing sample collection points

**Geology of the Study Area**

Eleyele is composed of a banded gneiss in which hornblende-biotite rich bands alternate with quartz, oligoclase rich bands (Figure B). The banded gneiss which orients as part of a sedimentary sequence contain a large sequence of gneiss and thin percolated layer of quartzite and amphibole. These rocks can be grouped into major and minor rock types. The major types are quartz of the meta-sedimentary series and the migmatite complex comprising banded gneiss, augen gneiss and magnetite, where the minor rock types include pegmatite, aplite, diorites, amphibolite and xenoliths (Akintola *et al*., 1994). The area has a low gentle undulating top composed primarily of banded gneiss in which hornblende- biotite rich bands alternate with quartz-plagioclase rich bands. The banded gneiss, which originated as part of a sedimentary sequence contains large lenses of granite gneiss and thin intercalated layers of quartzite and amphibolite.

Two distinct structural events can be identified in the early geological history of the Ibadan area; detailed field study of Ibadan area suggests that the formation of Ibadan granite gneiss which has yielded Eburnean Rb-Sr isochrones age was associated with the later of these events. Five phases of dyke end vein formation, two of which predate the formation of the granite gneiss have also been identified giving an overall sequence of geological events of which may correspond to the beginning of Liberian Orogeny around 3000MYA and the last of which the waning of the Pan Africa thermo tectonic event 5000 MYA (geology of Ibadan, wiki.org)

The type of rock in an area is an important factor governing the characteristics of its groundwater.

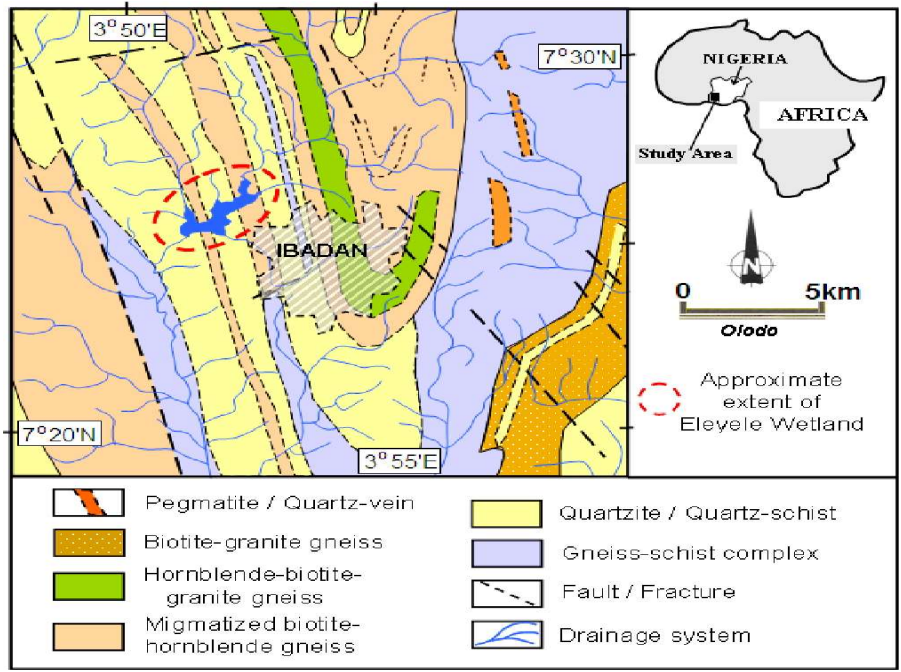

Figure 2: Geological map of the study area (Tijani *et al*, 2011)

## MATERIALS AND METHOD

The field work was carried out in Ibadan, Oyo state, Southwestern Nigeria. Twelve water samples were collected from Eleyele lake at different locations using a Geographic Positioning System (GPS) to determine the position of different locations. During the mapping, the sampling was carried out during the rainy season.

Some physical parameters of water were determined in-situ such as Total Dissolved Solids (TDS) using total dissolved solids meter, Electrical Conductivity (EC) using electrical conductivity meter, alkalinity and acidity of water using pH meter, water temperature and atmospheric temperature. Concentrated nitric acid ($HNO_3$) was added to the samples for cations in order to prevent precipitations of cations in the water sample before the water samples were analyzed for the various cations in the laboratory.

### Sampling Procedures

Sediment and water samples were collected from different locations in the stream; upstream, downstream and feeder stream of the lake.

1. All stream sediment sampling equipment were washed (bucket, polythene, gloves, plastic bottles for storing water)

2. Water samples of the same location were taken into two different plastic containers to determine the cation and anion which are 60ml and 1000ml respectively.

3. The samples taken for the cation were preserved using concentrated nitric acid to prevent oxidation of the metals which may in turn result in contamination

4. Sediment samples were sieved from dirt and rubbles and taken carefully into the polythene bags while wearing gloves.

5. Each sample is labelled according to the location number using a permanent marker board.

6. Each sample location is recorded with the coordinate using the GPS

## RESULTS AND DISCUSSIONS

### Result of physicochemical parameters and interpretation

The results of the physicochemical parameters of the water samples in the investigated study area are presented in Table 1.

### Total Dissolved Solids (TDS)

It can be observed that elements, minerals and salts are equally distributed in the main body of the lake except for certain places with slight variation compared to the feeder streams where there is great variation. The lake has a TDS value range from 114ppm at 21ML9 to 129ppm at 21ML11. Although there are certain places where the TDS value spikes, these values still meet the requirement value for TDS by WHO, which stated the maximum limit of TDS in quality water as 1200ppm (WHO, 2012). However, the TDS value of 1200ppm is regarded as poor but acceptable but pH greater than 1200ppm is largely unacceptable. The profile showing the variation of TDS is shown in Figure 3.

### Electrical Conductivity (EC)

The electrical conductivity is constant for the mainstream having a value of 100μs/cm all through the various locations. The maximum limit for EC value is 2500μs/cm at 20°C as stated by WHO (2012). The profile showing the variation of EC is shown in Fig 4.

**pH Concentration**

The pH values obtained varies from 6.9 to 7.5 as seen in table 1. It has an average value of 7.17 which is close to the value of neutral water (7), indicating that the water is non acidic. The pH limit for drinking water has been set to 6.5 to 8.5 as the standard for a quality water (WHO,2012). The profile showing the variation of pH is shown in Fig 5.

**Water Temperature (T$_W$)**

The temperature of the water body in the study area river has its values ranging from 26.5 °C to 28.2 °C. This is well below the maximum limit set by WHO which is 400 °C (WHO,2012).

**Atmospheric Temperature (T$_A$)**

The temperature of the atmosphere above the sample points on the Eleyele main lake river has its values ranging from 26.5°C to 31.7°C. The profile showing the variation of temperature is shown in Figure 6.

Other parameters which determine a quality water include colour, odour and taste, the water at the Eleyele lake was really filthy and looked to be almost brownish green in colour with vegetation spread out on its surface and aquatic bugs flitting from point to point on the surface. The waster did not look sanitary enough to attempt a taste and though special attention wasn't paid here, from inside the canoe, the water did not smell strange.

Table 1: Statistical Presentation of the Physicochemical Parameters of Eleyele Main Lake

| Sample ID | Coordinates | pH | TDS (ppm) | $T_W$ (ºC) | $T_A$ (ºC) | EC (μS/cm) |
|---|---|---|---|---|---|---|
| 21ML1 | N7º25'12" E3 º 50'54" | 7 | 119 | 26.5 | 27.2 | 100 |
| 21ML2 | N7 º 25'25.8" E3 º 51'45.3" | 7.4 | 118 | 27.2 | 28.3 | 100 |
| 21ML3 | N7 º 25'27.7" E3 º 51'45.8" | 7.1 | 118 | 27.3 | 27.2 | 100 |
| 21ML4 | N7 º 25'29.8" E3 º 51'43.6" | 7.3 | 122 | 27.2 | 28.2 | 100 |
| 21ML5 | N7 º 25'30.4" E3 º 51'40.4" | 7.1 | 127 | 27.2 | 27.2 | 100 |
| 21ML6 | N7 º 25'29.9" E3 º 51'36.6" | 7.1 | 118 | 27.2 | 27.3 | 100 |
| 21ML7 | N7 º 25'30" E3 º 51'31.2" | 7.3 | 126 | 27.2 | 28.1 | 100 |
| 21ML8 | N7 º 25'29.5" E3 º 51'22.3" | 7.2 | 126 | 26.9 | 29.9 | 100 |
| 21ML9 | N7 º 25'29.75" E3 º 51'13.9" | 7 | 114 | 26.8 | 26.5 | 100 |
| 21ML10 | N7 º 25'29.5" E3 º 51'14.9" | 7.1 | 125 | 27.2 | 31.7 | 100 |
| 21ML11 | N7 º 25'24.44" E3 º 51'20.1" | 6.9 | 129 | 28.2 | 31.7 | 100 |
| 21ML12 | N7 º 25'23" E3 º 51'29.4" | 7.5 | 124 | 27.9 | 29.9 | 100 |
| Average | - | 7.2 | 122.2 | 27.2 | 28.6 | 100 |
| Minimum | - | 6.9 | 114 | 26.5 | 26.5 | 100 |
| Maximum | - | 7.5 | 129 | 28.2 | 31.7 | 100 |
| Standard Deviation | - | 0.2 | 4.5 | 0.4 | 1.7 | 0 |
| Mode | - | 7.1 | 118 | 27.2 | 27.2 | 100 |

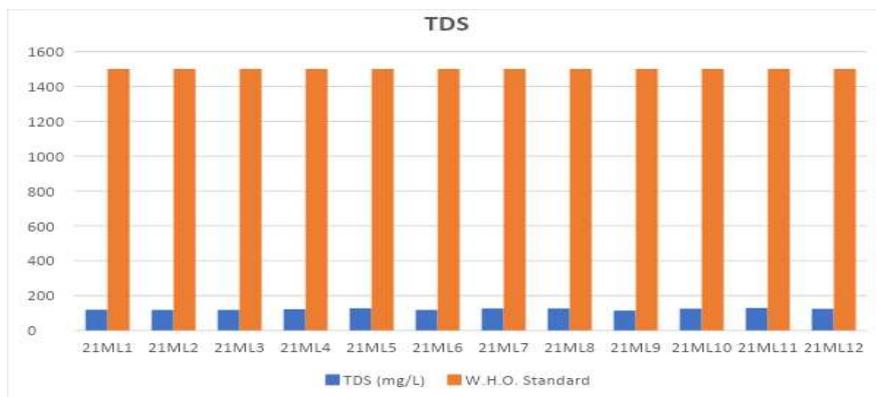

Figure 3: Profile showing the TDS variation the of the study area.

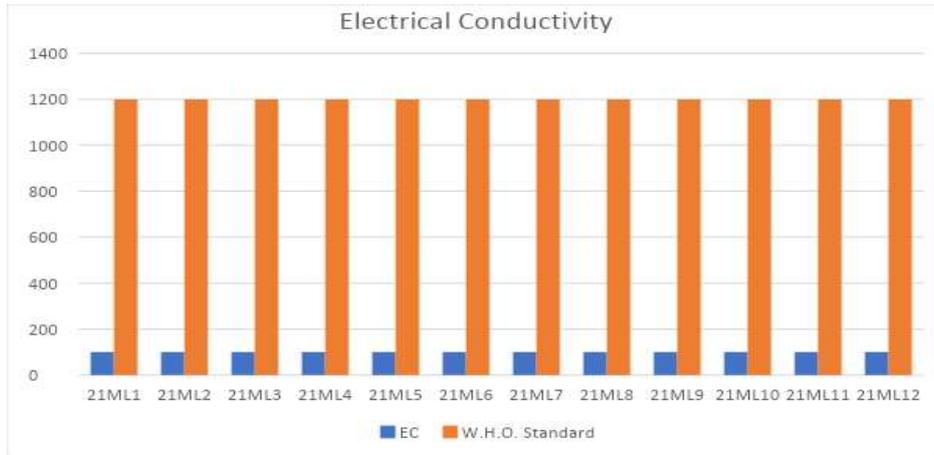

Figure 4: Profile showing the Electrical resistivity variation of the study area.

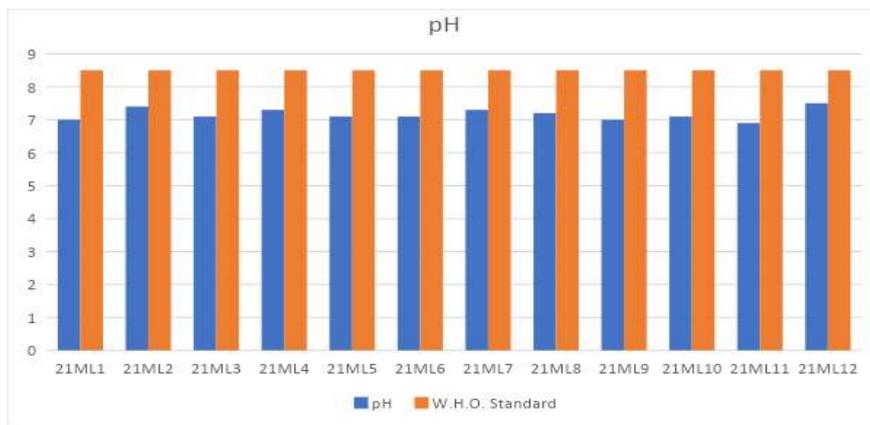

Figure 5: Profile showing the pH variation the of the study area.

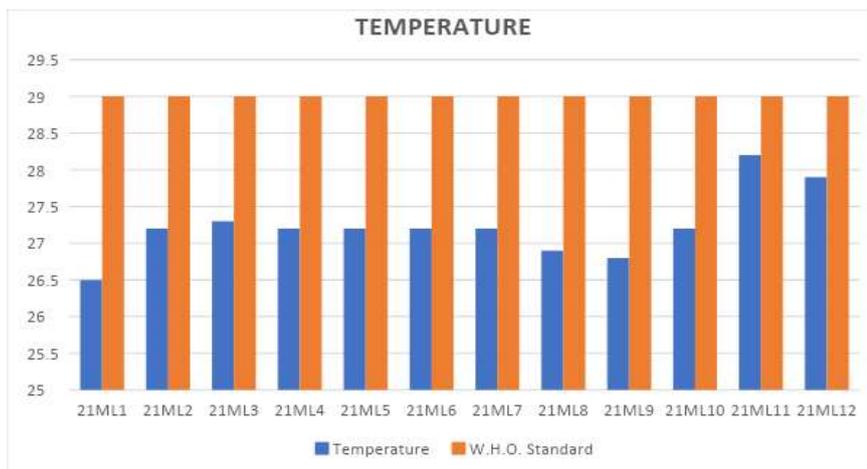

Figure 6: Profile showing the temperature variation the of the study area.

**Presentation and Interpretation of Trace Elements**

The different concentration in parts per billion (PPB) of the level of trace earth elements at various locations in Eleyele lake is presented in table 4.2(2a and 2b). The data shows variation in concentrations of these trace earth elements present in the study area. Though 12 samples were taken, only eight were analyzed because of financial constraints.

Elements like Ag, As, Be, Bi, Cd, Ga, Se, Sn, Hf, Nb, Ta, Tl, W, among others show very little variation in their results to be properly assessed and investigated. Some other elements have no exact value but an estimated value (<0.05 for example), those elements were also not included in my study for quality control reasons.

Table 2a: Presentation of Concentration of Trace Element present in the Study Area in PPB

| Location | Ag | Al | As | Ba | Be | Bi | Cd | Co | Cr | Cs | Cu | Fe | Ga | Hf |
|---|---|---|---|---|---|---|---|---|---|---|---|---|---|---|
| | PPB | PPB | PPB | PPB | PPB | PPB | PPB | PPB | PPB | PPB | PPB | PPB | PPB | PPB |
| MDL | 0.05 | 1 | 0.5 | 0.05 | 0.05 | 0.05 | 0.05 | 0.02 | 0.5 | 0.01 | 0.1 | 10 | 0.05 | 0.02 |
| 21ML01 | <0.05 | 123 | <0.5 | 108.41 | <0.05 | <0.05 | <0.05 | 1.24 | <0.5 | 0.04 | 2.7 | 656 | <0.05 | <0.02 |
| 21ML07 | <0.05 | 90 | <0.5 | 98.62 | <0.05 | <0.05 | <0.05 | 0.75 | <0.5 | 0.04 | 1.1 | 377 | <0.05 | <0.02 |
| 21ML10 | <0.05 | 104 | <0.5 | 102.42 | <0.05 | <0.05 | <0.05 | 0.8 | <0.5 | 0.04 | 2.2 | 408 | <0.05 | <0.02 |
| 21ML12 | <0.05 | 151 | <0.5 | 101.97 | <0.05 | <0.05 | <0.05 | 0.85 | 0.7 | 0.05 | 2.4 | 564 | 0.05 | <0.02 |
| 21ML02 | <0.05 | 25 | <0.5 | 47.64 | <0.05 | <0.05 | <0.05 | 0.43 | 1.1 | 0.04 | 0.8 | 635 | <0.05 | <0.02 |
| 21ML03 | <0.05 | 14 | <0.5 | 110.6 | <0.05 | <0.05 | <0.05 | 0.28 | 0.9 | 0.03 | 0.5 | 522 | <0.05 | <0.02 |
| 21ML04 | <0.05 | 27 | <0.5 | 102.31 | <0.05 | <0.05 | <0.05 | 0.3 | 1.2 | 0.04 | 0.8 | 590 | <0.05 | <0.02 |
| 21ML05 | <0.05 | 11 | <0.5 | 105.08 | <0.05 | <0.05 | <0.05 | 0.27 | 0.6 | 0.03 | 0.5 | 630 | <0.05 | <0.02 |

The trace elements present in the lake are very important in determining the quality of water of the river.in this chapter, the trace elements are properly discussed and are compared with the maximum permissible limit as determined by the World Health Organization.

The concentration of trace elements suitable for analysis are presented in Table 4.3 below.

Figure 7 displays these variations in a line chart. The visual representation indicates iron has the highest concentration in all the water samples (21ML1-5 and 21ML7,10 and 12).

Table 2b: Presentation of Concentration of Trace Element present in the Study Area in PPB (continued)

| Location | Mo | Nb | Rb | Sb | Se | Sn | Sr | Ta | Tl | U | V | W | Zn | Zr |
|---|---|---|---|---|---|---|---|---|---|---|---|---|---|---|
| | PPB | PPB | PPB | PPB | PPB | PPB | PPB | PPB | PPB | PPB | PPB | PPB | PPB | PPB |
| MDL | 0.1 | 0 | 0 | 0.05 | 0.5 | 0.05 | 0.01 | 0.02 | 0 | 0.02 | 0.2 | 0 | 0.5 | 0.02 |
| 21ML01 | 0.4 | <0.01 | 9.7 | 0.24 | <0.5 | 0.21 | 172 | <0.02 | 0 | 0.1 | 2.4 | <0.02 | 25.7 | 0.04 |
| 21ML07 | 0.4 | <0.01 | 9.4 | 0.15 | <0.5 | <0.05 | 168 | <0.02 | 0 | 0.09 | 1.7 | <0.02 | 6.9 | 0.03 |
| 21ML10 | 0.4 | <0.01 | 9.7 | 0.2 | <0.5 | <0.05 | 168 | <0.02 | 0 | 0.12 | 1.7 | <0.02 | 23.4 | 0.03 |
| 21ML12 | 0.4 | <0.01 | 9.9 | 0.19 | <0.5 | <0.05 | 167 | <0.02 | 0 | 0.12 | 1.9 | <0.02 | 63.1 | 0.03 |
| 21ML02 | 0.2 | <0.01 | 11 | <0.05 | <0.5 | <0.05 | 143 | <0.02 | <0.01 | <0.02 | 0.5 | <0.02 | 1.6 | 0.02 |
| 21ML03 | 0.2 | <0.01 | 11 | <0.05 | <0.5 | <0.05 | 140 | <0.02 | <0.01 | <0.02 | 0.3 | <0.02 | 0.7 | <0.02 |
| 21ML04 | 0.1 | <0.01 | 11 | 0.05 | <0.5 | <0.05 | 141 | <0.02 | <0.01 | <0.02 | 0.3 | <0.02 | 2 | <0.02 |
| 21ML05 | 0.1 | <0.01 | 11 | <0.05 | <0.5 | <0.05 | 142 | <0.02 | <0.01 | <0.02 | 0.3 | <0.02 | 0.9 | <0.02 |

Table 3: Concentration of Trace Element present in the Study Area in PPB, suitable for analysis.

| Location | Al | Ba | Co | Cs | Cu | Fe | Mo | Rb | Sr | V | Zn |
|---|---|---|---|---|---|---|---|---|---|---|---|
| 21ML01 | 123 | 108 | 1.2 | 0.04 | 2.7 | 656 | 0.4 | 9.74 | 172 | 2.4 | 26 |
| 21ML02 | 25 | 48 | 0.4 | 0.04 | 0.8 | 635 | 0.2 | 11 | 143 | 0.5 | 1.6 |
| 21ML03 | 14 | 111 | 0.3 | 0.03 | 0.5 | 522 | 0.2 | 11.2 | 140 | 0.3 | 0.7 |
| 21ML04 | 27 | 102 | 0.3 | 0.04 | 0.8 | 590 | 0.1 | 11.1 | 141 | 0.3 | 2 |
| 21ML05 | 11 | 105 | 0.3 | 0.03 | 0.5 | 630 | 0.1 | 11.4 | 142 | 0.3 | 0.9 |
| 21ML07 | 90 | 99 | 0.8 | 0.04 | 1.1 | 377 | 0.4 | 9.35 | 168 | 1.7 | 6.9 |
| 21ML10 | 104 | 102 | 0.8 | 0.04 | 2.2 | 408 | 0.4 | 9.66 | 168 | 1.7 | 23 |
| 21ML12 | 151 | 102 | 0.9 | 0.05 | 2.4 | 564 | 0.4 | 9.91 | 167 | 1.9 | 63 |
| **Minimum** | 11 | 48 | 0.3 | 0.03 | 0.5 | 377 | 0.1 | 9.35 | 140 | 0.3 | 0.7 |
| **Maximum** | 151 | 111 | 1.2 | 0.05 | 2.7 | 656 | 0.4 | 11.4 | 172 | 2.4 | 63 |
| **Average** | 68.1 | 97 | 0.6 | 0.04 | 1.38 | 547.8 | 0.28 | 10.4 | 155 | 1.14 | 16 |

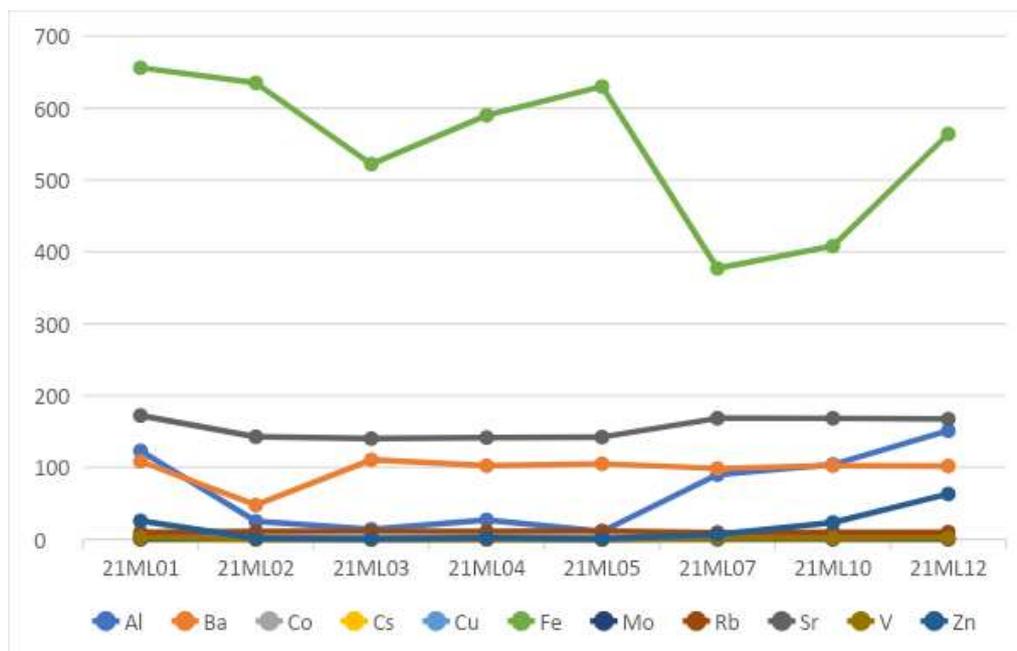

Figure 7: Profile showing variation in concentration of selected trace earth elements in Eleyele Lake.

**Aluminum Concentration (Al)**

The concentration of aluminum in Eleyele lake varies across the different sample collection points with an average value of 68.1ppb. Its lowest value is 11ppb recorded at 21ML05 while the highest value is 151, recorded at 21ML12 (Figure 8). This all falls within the range of drinkable water as recommended by the WHO guidelines which is between 100ppb – 200ppb (WHO, 2004). The main source of aluminum in the main lake is suspected to be geogenic, i.e. from rock and soil leaching.

**Barium Concentration (Ba)**

Barium's concentration in Eleyele lake varies across the different sample collection points with an average value of 97ppb. Its lowest value is 48ppb recorded at 21ML02 while the highest value is 111, recorded at 21ML03 (Figure 9). This all falls well short of the Maximum Contaminant Level (MCL) as recommended by the WHO guidelines which is 700ppb (WHO, 2004). The main source of barium in the main lake in this quantity is suspected to be geogenic.

**Cobalt Concentration (Co)**

The concentration of cobalt in Eleyele lake varies across the different sample collection points with an average value of 0.6ppb. Its lowest value is 0.3ppb which was recorded at three points, 21ML03, 21ML04 and 21ML05. The highest value is 1.2ppb, recorded at 21ML01, this falls within the range acceptable for drinking water which is 1-2ppb (Figure 10). Overall, the cobalt levels are low in the Eleyele lake.

**Cesium Concentration (Cs)**

The concentration of cesium in Eleyele lake is relatively low and the values show very little variation across collection points. The lowest value recorded was 0.3ppb at 21ML03 and 21ML05. The average value was 0.4ppb and it was recorded five times at 21ML01, 21ML02, 21ML04, 21ML07 and 21ML10. The highest value, 0.5ppb, was recorded at 21ML12 (Figure 11). Cesium's values were the lowest recorded among all the trace elements across the board.

**Copper Concentration (Cu)**

The concentration of copper in Eleyele lake varies across the different sample collection points with an average value of 1.38ppb. Its lowest value is 0.5ppb recorded at two locations, 21ML03 and 21ML05, while the highest value is 2.7ppb, recorded at 21ML01. (Figure 12).

The concentration of copper safe for consuming has a limit of 2 ppm (2000ppb) as specified by WHO (WHO, 2011) in quality water.

**Iron Concentration (Fe)**

The concentration of iron in Eleyele lake varies across the different sample collection points with an average value of 547.8ppb. Its lowest value is 377ppb, recorded at 21ML07, while the highest value is 656ppb, recorded at 21ML01 (Figure 13). Iron's values are the highest set of values recorded among all the trace elements across the board.

This all falls outside the limit of concentration given by WHO which is 300ppb (WHO 2011). The main source of iron in the main lake in this quantity is suspected to be anthropogenic.

**Molybdenum Concentration (Mo)**

The concentration of molybdenum in Eleyele lake varies little across the sample collection points with an average value of 0.28ppb. Its lowest value, 0.1ppb, was recorded at points 21ML04 and 21ML05, while the highest value, 0.4ppb, was recorded at 21ML01, 21ML07, 21ML10 and 21ML12 (Figure 14). Molybdenum's values are the second lowest set of values recorded among all the trace elements across the board, second only to cesium.

This all falls well short of the WHO guideline value of 20ppb. The main source of barium in the main lake in this quantity is suspected to be geogenic.

**Rubidium Concentration (Rb)**

The concentration of rubidium in Eleyele lake varies across the sample collection points with an average value of 10.4ppb. Its lowest value, 9.35ppb, was recorded at 21ML07, while the highest value, 11.4ppb, was recorded at 21ML05(Figure 15).

This all falls well short of the Maximum Contaminant Level (MCL) as recommended by the WHO guidelines which is 50ppb. The main source of barium in the main lake in this quantity is suspected to be geogenic.

**Strontium Concentration (Sr)**

The concentration of strontium in Eleyele lake varies across the sample collection points with an average value of 155ppb. Its lowest value, 140ppb, was recorded at 21ML03, while the highest value, 172ppb, was recorded at 21ML01(Figure 16). Strontium's values are the second highest set of values recorded among all the trace elements across the board, only iron had higher values recorded.

**Vanadium Concentration (V)**

The concentration of vanadium in Eleyele lake varies across the sample collection points with an average value of 1.14ppb. Its lowest value, 0.3ppb, was recorded at points 21ML03, 21ML04 and 21ML05, while the highest value, 2.4ppb, was recorded at 21ML01(Figure 17).

The concentration of vanadium in water is largely dependent on geographical location and ranges from 0.2 to more than 100ppb in freshwater.

**Zinc Concentration (Zn)**

The concentration of zinc in Eleyele lake varies greatly across the sample collection points with an average value of 16ppb. Its lowest value, 0.7ppb, was recorded at 21ML03, while the highest value, 63ppb, was recorded at 21ML12 (Figure 18).

This all falls well short of the Maximum Contaminant Level (MCL) as recommended by the WHO guidelines which is 400ppb (WHO, 2004). The main source of zinc in the main lake in this quantity is suspected to be geogenic.

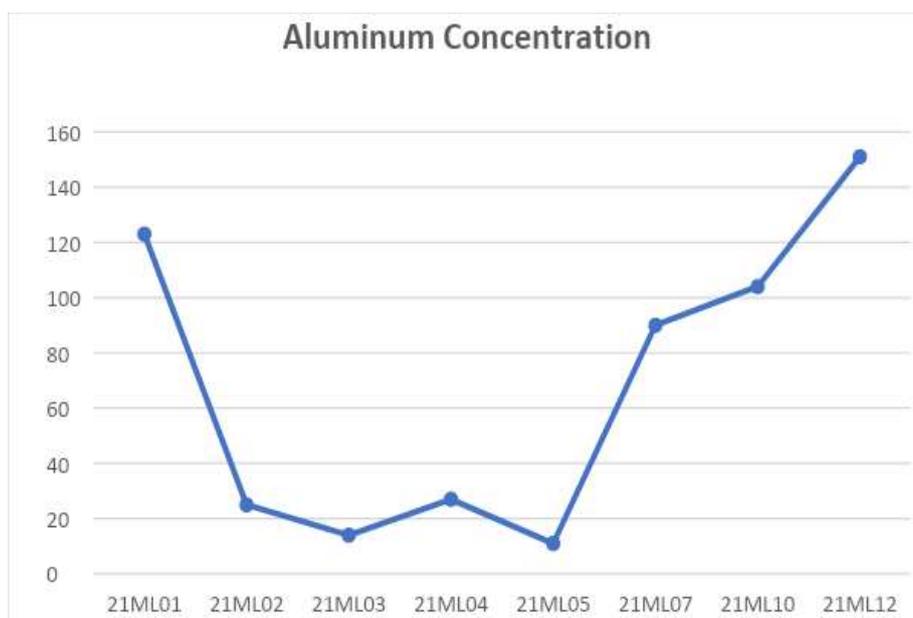

Figure 8; Variation of Aluminum Concentration in Eleyele Lake

Kunle-John, I. O., Micheals, S. P., and Okay, E. N.
Water Resources Vol. 34 No.2 (2024) 190 – 210

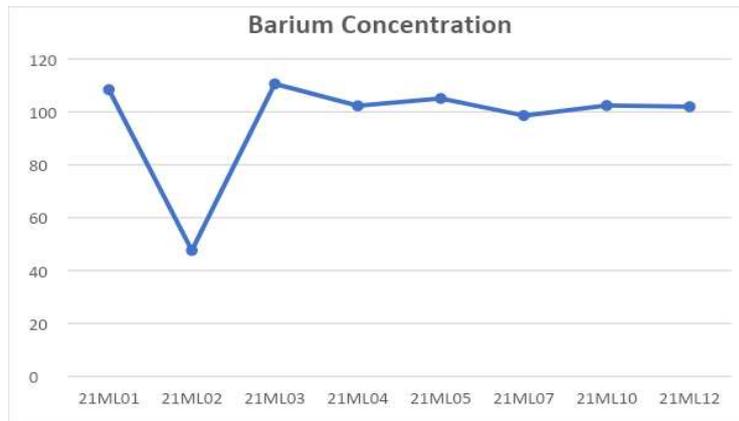

Figure 9; Variation of Barium Concentration in Eleyele Lake

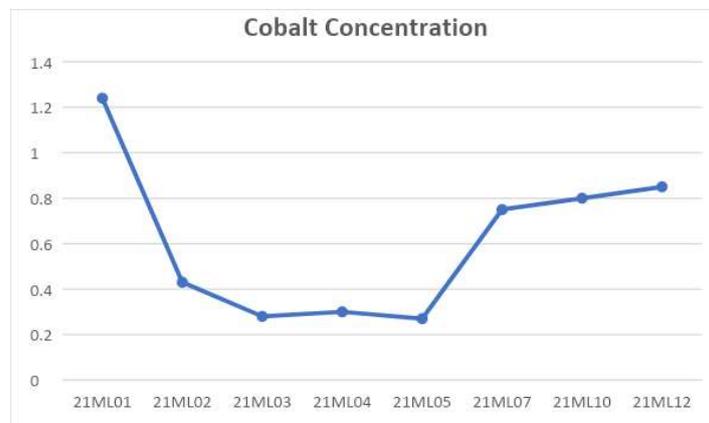

Figure 10; Variation of Cobalt Concentration in Eleyele Lake

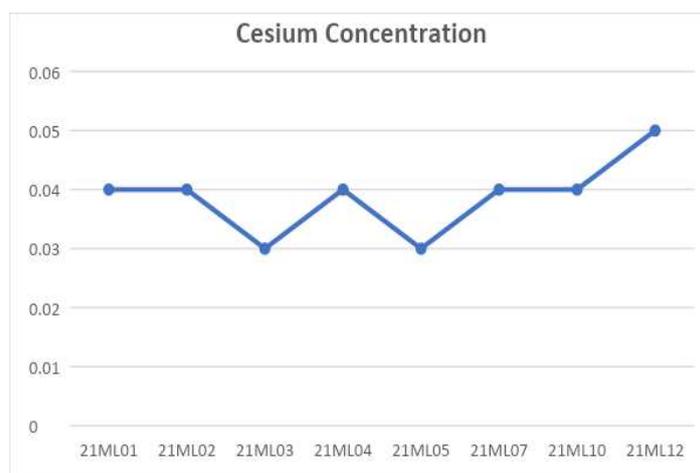

Figure 11: Variation of Cesium Concentration in Eleyele Lake.

Kunle-John, I. O., Micheals, S. P., and Okay, E. N.
Water Resources Vol. 34 No.2 (2024) 190 – 210

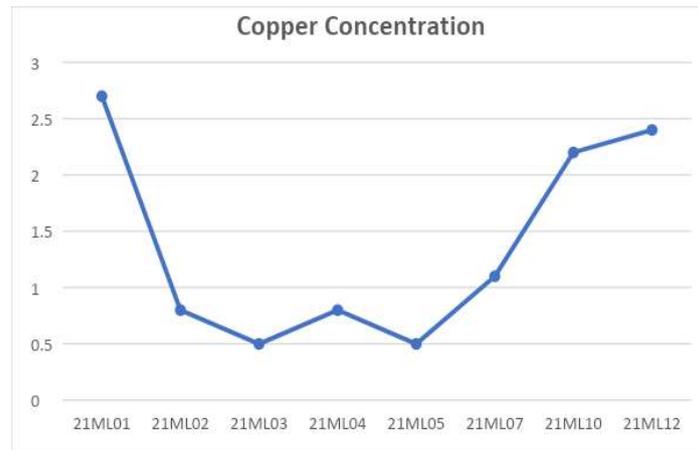

Figure 12: Variation of Copper Concentration in Eleyele Lake

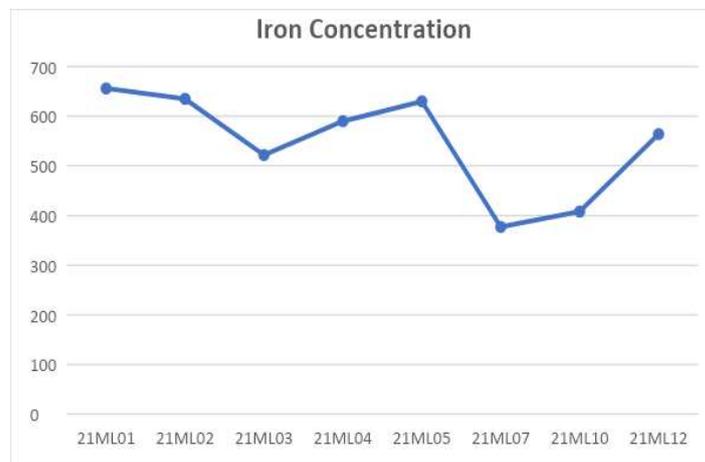

Figure 13: Variation of Iron Concentration in Eleyele Lake

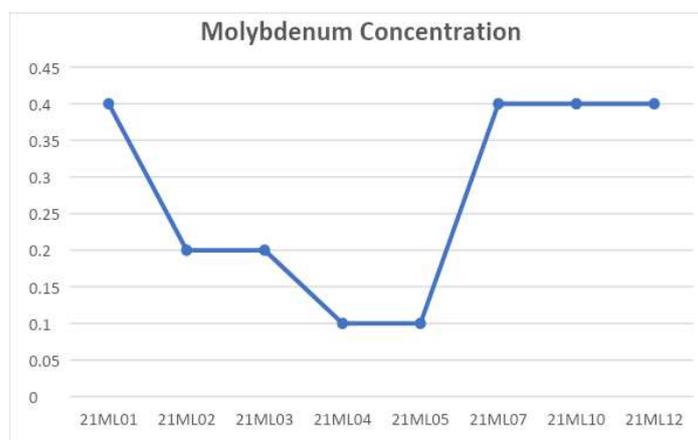

Figure 14: Variation of Molybdenum Concentration in Eleyele Lake

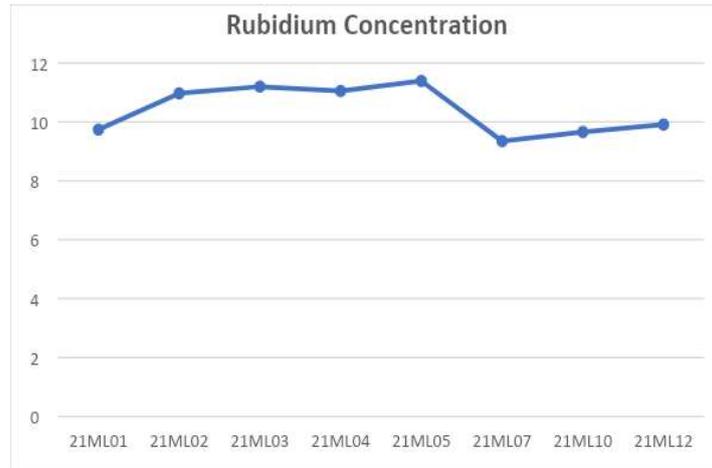

Figure 15: Variation of Rubidium Concentration in Eleyele Lake

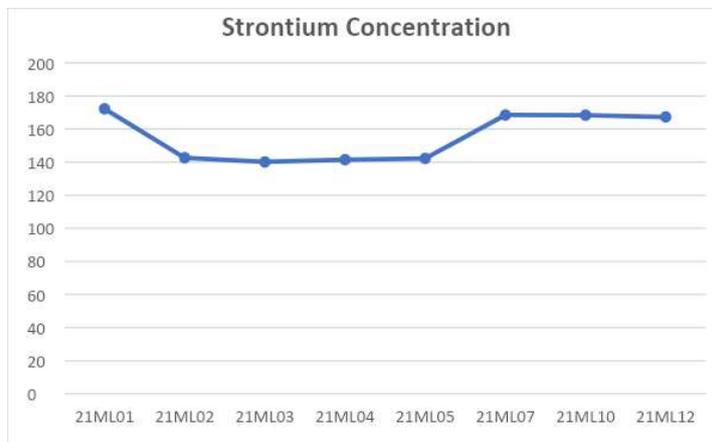

Figure 16: Variation of Strontium Concentration in Eleyele Lake

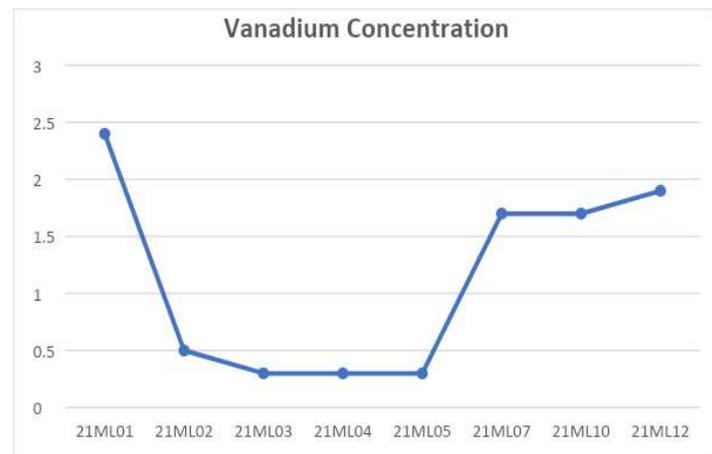

Figure 17: Variation of Vanadium Concentration in Eleyele Lake

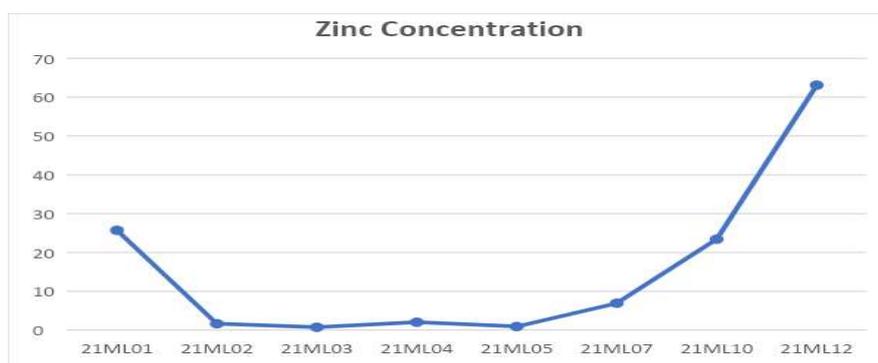

Figure 18: Variation of Zinc Concentration in Eleyele Lake

**Assessment of Contamination**

The assessment of contamination describes the level of contamination of elements in water. The following methods are used to determine the extent of contamination in the water samples collected from Eleyele lake.

**Contamination Factor**

The contamination factor has been calculated for each location in Eleyele lake to determine the amount at which each element is present as contaminant in the river.

Table 4 shows the values for contamination factor of the trace earth elements in Eleyele lake. Table 5 shows the interpretation of contamination factors in Eleyele lake. Another source of interpretation according to Tijani *et al*. (2007) of these trace earth elements are presented. This source states that if the value of the contamination factor is less than 1 (cf<1), then the source is suggested to be geogenic (natural) and if it is greater than 1 (cf>1), it is suggested to be anthropogenic sources (human activities).

**Degree of Contamination**

The degree of contamination is the sum of the contamination factor values of the trace earth elements in the water samples from the different locations in the study area. Table 6 shows the average degree of contamination of trace earth elements in Eleyele lake and their respective interpretations of these rare earth elements where the highest level of degree of contamination is in sample 21ML12 with value of 0.134, though rare earth elements are known to be very minute in surface water, but the high value of the degree of contamination signify the extent of which the water is being contaminated with rare earth elements.

**Geo-Accumulation Index**

The results of the geo-accumulation index of nine water samples in Eleyele lake are presented in table 4. Table 4 shows the interpretation of contamination factor of the average values of all the rare earth elements.

Table 4: Contamination factor of the level of Trace earth elements in Eleyele lake

| Location | Al | Ba | Co | Cs | Cu | Fe | Mo | Rb | Sr | V | Zn |
|---|---|---|---|---|---|---|---|---|---|---|---|
| 21ML01 | 14.1 | 0.21 | 0.06 | 0.0064 | 0.036 | 112.9 | 0.13 | 0.124 | 0.921 | 0.019 | 0.124 |
| 21ML02 | 2.87 | 0.09 | 0.02 | 0.0064 | 0.011 | 109.3 | 0.07 | 0.14 | 0.763 | 0.004 | 0.008 |
| 21ML03 | 1.61 | 0.21 | 0.01 | 0.0048 | 0.007 | 89.85 | 0.07 | 0.143 | 0.749 | 0.002 | 0.003 |
| 21ML04 | 3.1 | 0.2 | 0.01 | 0.0064 | 0.011 | 101.5 | 0.03 | 0.141 | 0.757 | 0.002 | 0.01 |
| 21ML05 | 1.26 | 0.2 | 0.01 | 0.0048 | 0.007 | 108.4 | 0.03 | 0.145 | 0.761 | 0.002 | 0.004 |
| 21ML07 | 10.3 | 0.19 | 0.03 | 0.0064 | 0.014 | 64.89 | 0.13 | 0.119 | 0.901 | 0.013 | 0.033 |
| 21ML10 | 11.9 | 0.2 | 0.04 | 0.0064 | 0.029 | 70.22 | 0.13 | 0.123 | 0.9 | 0.013 | 0.113 |
| 21ML12 | 17.3 | 0.2 | 0.04 | 0.008 | 0.032 | 97.07 | 0.13 | 0.126 | 0.895 | 0.015 | 0.303 |
| Average | 7.81 | 0.19 | 0.03 | 0.0062 | 0.018 | 94.28 | 0.09 | 0.133 | 0.831 | 0.009 | 0.075 |
| Minimum | 1.61 | 0.09 | 0.01 | 0.0048 | 0.007 | 64.89 | 0.03 | 0.119 | 0.749 | 0.002 | 0.003 |
| Maximum | 17.3 | 0.21 | 0.06 | 0.008 | 0.036 | 112.9 | 0.13 | 0.145 | 0.921 | 0.019 | 0.303 |
| Standard deviation | 5.9 | 0.04 | 0.02 | 0.00096 | 0.01 | 16.9 | 0.04 | 0.0099 | 0.074 | 0.0065 | 0.098 |
| Mode | - | 0.2 | 0.01 | 0.0064 | - | - | 0.13 | - | - | 0.002 | - |

Table 5: Interpretation of the average value of contamination factor of all trace earth elements in Eleyele lake.

| Trace Earth elements | Average Contamination Factor | Interpretation |
|---|---|---|
| Aluminum | 7.8125 | High degree of contamination |
| Barium | 0.186075192 | Low degree of contamination |
| Cobalt | 0.027333333 | Low degree of contamination |
| Cesium | 0.0062 | Low degree of contamination |
| Copper | 0.018115942 | Low degree of contamination |
| Iron | 94.27710843 | High degree of contamination |
| Molybdenum | 0.0923 | Low degree of contamination |
| Rubidium | 0.132595541 | Low degree of contamination |
| Strontium | 0.830715241 | Low degree of contamination |
| Vanadium | 0.008817829 | Low degree of contamination |
| Zinc | 0.074699519 | Low degree of contamination |
| Average | 9.40604 | - |
| Minimum | 0.0062 | - |
| Maximum | 94.27710843 | - |
| Standard deviation | 26.93 | - |

Table 6: Degree of Contamination and interpretation of the trace earth elements in Eleyele lake

| Sample number | Degree of contamination | Interpretation |
|---|---|---|
| 21ML01 | 129 | High degree of contamination |
| 21ML02 | 113 | High degree of contamination |
| 21ML03 | 93 | High degree of contamination |
| 21ML04 | 106 | High degree of contamination |
| 21ML05 | 111 | High degree of contamination |
| 21ML07 | 77 | High degree of contamination |
| 21ML10 | 84 | High degree of contamination |
| 21ML12 | 116 | High degree of contamination |
| Average | 103.63 | - |
| Minimum | 77 | - |
| Maximum | 129 | - |
| Standard deviation | 16.4 | - |

Table 7: Geo-Accumulation index of the level of trace earth elements in Eleyele lake

| Location | Al | Ba | Co | Cs | Cu | Fe | Mo | Rb | Sr | V | Zn |
|---|---|---|---|---|---|---|---|---|---|---|---|
| 21ML01 | 3.23 | -2.85 | -4.8 | -7.87 | -5.4 | 6.23 | -3.48 | -3.6 | -0.7 | -6.33 | -3.6 |
| 21ML02 | 0.93 | -4.04 | -6.3 | -7.87 | -7.15 | 6.19 | -4.48 | -3.42 | -0.98 | -8.6 | -7.61 |
| 21ML03 | 0.1 | -2.82 | -6.9 | -8.29 | -7.83 | 5.9 | -4.48 | -3.39 | -1 | -9.33 | -8.8 |
| 21ML04 | 1.05 | -2.94 | -6.8 | -7.87 | -7.15 | 6.08 | -5.48 | -3.41 | -0.99 | -9.33 | -7.29 |
| 21ML05 | -0.2 | -2.9 | -7 | -8.29 | -7.83 | 6.18 | -5.48 | -3.37 | -0.98 | -9.33 | -8.44 |
| 21ML07 | 2.78 | -2.99 | -5.5 | -7.87 | -6.69 | 5.43 | -3.48 | -3.65 | -0.74 | -6.83 | -5.5 |
| 21ML10 | 2.99 | -2.93 | -5.4 | -7.87 | -5.69 | 5.55 | -3.48 | -3.61 | -0.74 | -6.83 | -3.74 |
| 21ML12 | 3.53 | -2.94 | -5.3 | -7.55 | -5.57 | 6.02 | -3.48 | -3.57 | -0.75 | -6.67 | -2.31 |
| Average | 1.8 | -3.05 | -6 | -7.94 | -6.67 | 5.95 | -4.23 | -3.5 | -0.86 | -7.91 | -5.91 |
| Minimum | -0.2 | -4.04 | -7 | -8.29 | -7.83 | 5.43 | -5.48 | -3.65 | -1 | -9.33 | -8.44 |
| Maximum | 3.53 | -2.82 | -4.8 | -7.55 | -5.4 | 6.23 | -3.48 | -3.37 | -0.7 | -6.33 | -2.31 |
| Standard Deviation | 1.4 | 0.38 | 0.8 | 0.23 | 0.93 | 0.28 | 0.83 | 0.11 | 0.13 | 1.27 | 2.31 |
| Mode | - | - | - | -7.87 | -7.83 | - | -3.48 | - | - | -9.33 | - |

Table 8: Interpretation of average geo-accumulation index of the trace earth elements in Eleyele lake

| Trace earth element | Geo-accumulation indices | Interpretation |
|---|---|---|
| Aluminum | 1.8 | Moderately unpolluted |
| Barium | -3.05 | Practically unpolluted |
| Cobalt | -6 | Practically unpolluted |
| Cesium | -7.94 | Practically unpolluted |
| Copper | -6.67 | Practically unpolluted |
| Iron | 5.95 | Extreme Pollution |
| Molybdenum | -4.23 | Practically unpolluted |
| Rubidium | -3.5 | Practically unpolluted |
| Strontium | -0.86 | Practically unpolluted |
| Vanadium | -7.91 | Practically unpolluted |
| Zinc | -5.91 | Practically unpolluted |
| Average | -3.48 | - |
| Minimum | -7.94 | - |
| Maximum | 5.95 | - |
| Standard Deviation | 4.12 | - |

**Environment Impact**

The environment is a constant in the life of humans. It provides sustenance among several benefits, but it can also harm individuals. Humans are responsible for their environment as our interactions with the environment determine its nature. Pollution by several careless means including poor waste disposal and human excreta is just some of the factors that negatively affect the environment and alter the natural state of its properties.

Trace elements exist widely in specific concentrations in the natural environment, with development of the economy and society of human activities, such as mining and smelting, the processing has allowed more trace elements to enter the atmosphere, water, and soil, thus resulting in serious environmental pollution. Pollution from trace elements has become harmful not only to the ecosystems, but also poses a threat to human health because of refractory characteristics of bioaccumulation. Trace elements are critical for life processes and sustainability, they are only needed at the trace level. Excess intake of essential trace elements in drinking water may lead to adverse health.

The trace elements polluting Eleyele lake in this study, iron and aluminum, have adverse effects on the life in the water body, soil and in humans too. Aluminum intake in high quantities causes adverse effects of intake of aluminum causes Osteoporosis which means regression of bone growth, vitamin D-resistant osteomalacia, erythropoietin-resistant microcytic anemia, and central nervous system, Iron also has it adverse effects in high quantities which causes serious conditions such as diabetes, heart problems and liver disease.

**CONCLUSION**

In conclusion, the results of physicochemical parameters show that TDS ranges from 114ppm to 129ppm with an average of 122.2ppm and the EC was uniform throughout various points of reading at 100μs/cm which suggests that the lake is a fresh water body which is suitable for drinking and other specific purposes. The pH readings range from 6.9 to 7.5 with an average of 7.2 which means the water is neither acidic nor basic but best described as neutral. Water temperature ranged from 26.5°C to 28.2°C with an average of 27.2°C which means that the water temperature is warm, and the temperature is not adversely affecting the water, since high temperature negatively impact water quality by enhancing the growth of microorganisms which may increase taste, odour, colour and corrosion problem. The concentration of the trace elements present in the water shows that the selected trace element falls within the WHO permissible limit standard. Also, the contamination indices of the selected elements for analysis showed that that Ba, Co, Cs, Cu, Mo, Rb, Sr, V, and Zn are generally less than 1 which means they are from geogenic source depicting their origin to be from the weathering of basement complex rock in the area, while Al and Fe are generally greater than 1 which means they are from anthropogenic source. The water has a high degree of contamination which is influenced by the high concentration of Aluminum and Iron which can be as a result of human activities and industrial waste disposal. It can be concluded from the result of the analysis that iron and aluminum have the highest concentration in Eleyele lake. Although aluminum and iron are essential elements for body growth and development, drinking water containing high levels of aluminum and iron can lead to anemia, osteomalacia (brittle or soft bones), cardiac arrest, stomach problems, nausea, and hemochromatosis. Iron in drinking water causes it to develop an unpleasant metallic taste.

Due to the high concentration of these two trace elements and the ensuing health risks, the Eleyele lake is not advisable for human consumption.

**Recommendations**

Based on this study, I hereby make these recommendations

1. A seasonal sampling should be done during the dry season to ascertain the actual concentration of elements in surface water and stream sediments before it is diluted by rainfall.
2. Government and local industries should focus on the development of cost-effective remediation and disposal techniques. These techniques should be taught to the residents, and they should also be sensitized on the dangers or careless waste disposal. Issues connected with deposition of the trace elements and environmental contaminants will clearly be resolved by reducing emissions and dumping of pollutants in the water.
3. The old environmental impact remediation methods are undergoing

further development, and it is necessary to continue to seek effective, economically affordable and environmentally sensitive techniques. Plans should be put in place for water treatment by the government or charitable non-governmental organizations in order for it to be suitable for the environment.

4. Microbial investigation should also be executed in other to determine the microbial action in the water in relation in relation to the role of waste dumps by the people living in the area.

5. Research on pathways, modes of exposure, and process mechanisms must continue in order to detect and develop important new techniques in environmental science. Also, the Government should conduct extensive studies on the geology of the country and release clear regulatory information on the limits of contaminants in the water bodies in the country. These limits should exist for both health studies (drinking water) and environmental studies (fresh water bodies). This will make the work of researchers easier as there will be ready access to the data.